\input{aipcheck}

\documentclass[
    ,final            % use final for the camera ready runs
  ]
  {aipproc}

\layoutstyle{6x9}

\begin{document}

\title{Relativistic Positioning Systems}


\author{Bartolomé COLL}{address={Syst\`emes de r\'ef\'erence
    relativistes, SYRTE-CNRS, Observatoire de Paris, 75014 Paris,
    France\\ bartolome.coll@obspm.fr}}

\begin{abstract}
The theory of relativistic {\em location systems} is sketched. An
interesting class of these systems is that of relativistic {\em
positioning systems,} which consists in sets of four clocks
broadcasting their proper time. Among them, the more important ones
are the {\em auto-located positioning systems,} in which every clock
broadcasts not only its proper time but the proper times that it
receives from the other three. At this level, no reference to any
exterior system (the Earth surface, for example) and no
synchronization are needed. Some properties are presented. In the
SYPOR project, such a structure is proposed,  eventually  anchored
to a classical reference system on the Earth surface, as the best
relativistic structure for Global Navigation Satellite Systems.
\end{abstract}

\maketitle

%%%%%%%%%%%%%%%%%%%%%%%%%%%%%%%%%%%%%%%%%%%%
%% MAINMATTER
%%%%%%%%%%%%%%%%%%%%%%%%%%%%%%%%%%%%%%%%%%%%

\section{Introduction}

In relativity, the physical space-time is modeled by a four-dimensional
differential manifold. So, it admits, in general, an infinite variety
of mathematical coordinate systems.

Among these abundant coordinate systems, only a few number of them is
known which may be physically {\em  interpreted}. This means that
(some of) their ingredients, namely their coordinate lines or
their coordinate (hyper-)surfaces, may be {\em imagined} as described
by some physical objects, like point-like particles or clocks, dust,
stretched strings or light signals.

Among these few physically interpretable coordinate systems,
only one of them is known which, generically, may be physically
{\em constructed}. This is the one based on the Poincaré-Einstein
synchronization procedure, i.e. by means of two-way signals,
sent by one observer equipped with a clock and returned by the events
he want to locate. This system, of the observer's clock and the two-way
signals, with the help of a theodolite, generates a four-dimensional
coordinate system with spatial spherical coordinates around the observer.
And this is the sole reasonable relativistic coordinate system that, up
to now, one has been able to construct physically in generic, arbitrarily chosen, vacuum space-times. It is also often called {\em radar system.}

But this relativistic physical coordinate system  suffers from an
important default: the one of being {\em intrinsically
retarded}. This means that the coordinates of every event in the (finite)
neighborhood of the observer are necessarily known with an unavoidable
delay not only by the observer, which, being separated from the event, expects such a delay,  but also by the event itself which is constitutively present at the instant and place where it happens and its coordinates indicate.

Consequently, up for the very particular circumstances in which the observer, the events and the whole gravitational context are stationary, even the events are unable to know their proper coordinates in this system. So, in it, the physical properties of an event cannot be {\em experimentally} related to its position, still less without delay; in these coordinates, such relations between properties and positions cannot but be {\em calculated} and need for this purpose the help of a previous theory (often unknown) of their proper evolution.

Thus, the main problem is {\em how to construct physically good, not intrinsically retarded, coordinate systems}, i.e. systems such that every event in it be able to know its proper coordinates without delay. The class of such {\em positioning systems} is relatively restricted, and their paradigmatic representatives are systems of four clocks broadcasting their proper time.

Our purpose here is to introduce the basic concepts, to comment them and to present some qualitative aspects. This is organized as follows: in Section {\sc location systems,} some general notions and properties concerning the physical realizations of coordinate systems are explained; in Section {\sc positioning systems} these systems are described and their principal properties presented, in particular the one of having a very good separation power for the space-time, and finally, in Section {\sc sypor project,} we shortly describe how these relativistic positioning systems should be used for primary reference and positioning of the Earth surroundings, replacing the at present Newtonian-relativistically-corrected conception of the Global Navigation Satellite Systems.

 Details on these and other results are presented in this meeting by my collaborators  \cite{Fe:2005}, \cite{Mo:2005}, \cite{Po:2005}.

\section{Location systems}

A coordinate system may be given in many different ways. But whatever they be, they are tantamount to give its (parameterized congruences of) {\em coordinate lines} or its (one-parameter families of) {\em coordinate (hyper)surfaces.} But lines and (hyper)surfaces may be physically constructed with many different materials and with many different protocols, giving raise to very different physical realizations of the same mathematical coordinate system. For this reason, it is convenient to distinguish by a different appellation coordinate systems ant their physical realizations. We call {\em location systems} the physical realizations of coordinate systems.

As physical realizations, location systems are physical objects and, consequently, able to be described in physical terms. For our purposes, the following physical description is sufficiently complete.

\begin{itemize}

\item %{\bf Definition {\rm(location system)}:}
{\em In a region of the space-time, a {\em location system} is a {\em real} or {\em virtual,} {\em passive,} set of {\em physical fields,} {\em parameterized} in such a way that every event in the region be one-to-one characterized by the values of the parameters at the event.}

\end{itemize}

In this physical description, 'real'  refers to the beforehand actual physical construction in all the domain of the whole set of physical fields\footnote{For example, the Cartesian  lines on a  graph paper, at the millimetric scale.}, meanwhile 'virtual' refers to any other case; in particular when  only the reference axes or surfaces are beforehand constructed, leaving afterward the construction of the sole lines or surfaces that contain the specific events of interest\footnote{For example,  the Cartesian system defined on a white sheet of paper by two orthogonal lines is virtual,  all the other lines of the congruence covering the sheet being not drawn beforehand and only those lines crossing the points of interest being afterward constructed.}.

By 'passive' set of fields  it is to be understood sufficiently weak physical fields so that their interaction with the events to be located may be considered negligible (a rigourous quantum field version of location systems would be necessarily 'active').

Finally, that a physical field is parameterized means that from the measure of some of its physical properties at every event of the domain, it is possible to extract a unique real number.

This description may be as well considered as the {\em  physical definition} of location systems, so that alternatively, one can consider either locations systems as physical realizations of coordinate systems or coordinate systems as mathematical idealizations of location systems.

The use of location systems may respond to different needs or objectives; two of them are particularly important. As in astronomy, frequently the goal of some location systems  is to allow one observer, generally considered at the origin, to locate with precision the events of his neighborhood. Location systems devoted to such a function are called ({\em relativistic}) {\em reference systems.} The goal of other location systems, like Global Navigation Satellite Systems (GNSS), is to indicate to every event of the region its own position. Location systems devoted to such a function are called ({\em relativistic}) {\em positioning systems.}\footnote{Reference and positioning systems defined here are {\em four-dimensional} objects, including time location. This is not still the common use, and so, the International Astronomical Union (IAU) considers separately time scales and (three-dimensional) reference systems. We believe that, from a relativistic point of view, it is imperative to gather them in a sole four-dimensional concept, if we want to adequate our points of view to the increasing and pressing presence of relativistic corrections. Also, the International Celestial Reference System (ICRS), in spite of its appellation, {\em is not} a reference system even in the three-dimensional sense, but only an {\em orientation system}; if at first glance on could consider these features as a simple matter of words, they induce to confusion students and professionals, delaying the construction of correctly conceived relativistic frames.}

In Newtonian theory, as far as the velocity of information is supposed to be infinite, both goals are exchangeable for any location system. But in general relativity this is no longer possible, and the goal of a location system strongly conditions its conception and its construction. In fact, one has a strong hierarchy between them: meanwhile it is impossible to construct a positioning system starting from a reference system by transmission of its data, it is always possible, and very easily, to construct a reference system starting from a positioning system (it is sufficient that every event send its coordinates to the observer). It is then evident that, whenever possible, it is a positioning system, and not a reference system, that has the most interest to be constructed. Of course, this it not always possible, as is the case, roughly speaking,  for the space out of the Solar system\footnote{A positioning system for the Solar system based on the signals of (basically) four millisecond pulsars has been proposed in \cite{CT}.}, but for such regions, people conforms with slightly more than an orientation system\footnote{See the end of next to last footnote.}, which is far from being the physical realization of a coordinate system (an orientation system is, in fact, nothing but a basis of the {\em tangent frame} to the observer for the light directions converging to him).

\section{Positioning systems}

Positioning systems are here supposed to be {\em generic,} {\em free} and  {\em immediate.}

A location system is {\em generic} (for a given class of space-times) if it can be constructed in any space-time (of the class). For example, Cartesian systems are not generic but for (the class of) Minkowski space-time, meanwhile harmonic systems are generic for (the class of) {\em all} space-times.

A location system is (gravity) {\em free} if its construction does not need the previous knowledge of the gravitational field\footnote{A location system is a physical object that lives in the physical space-time. In it, even if we do not know the metric, such objects as test particles, light rays or signals follow specific paths which, a priori, may allow constructing a location system.}. For example, harmonic systems are not free; in fact, among the usual location systems, only the radar system, is free.

A location system is {\em immediate} if every event of its domain may know its coordinates without delay. Immediate systems\footnote{From the Late Latin 'immediatus', 'without anything between'.} are the antithesis of the already mentioned intrinsically retarded systems, to which belongs the radar system. From the relativistic point of view, no one of the location systems known up to now are immediate.

The question is then if whether or not generic, free and immediate positioning systems can be constructed. Because involving real objects, the answer to this question is an {\em epistemic} answer, rather than a logical one, resulting from the analysis of at present methods, techniques and practical possibilities of physical construction of such systems. This analysis show that the set of generic, free and immediate relativistic positioning  systems constitute a small class of location systems. As already mentioned, the paradigmatic representatives of this class are the location systems constituted by four clocks broadcasting their proper time.

\vspace{0.5cm}

In what sense four clocks broadcasting their proper time constitute a physical realization of a coordinate system? A coordinate system is defined by its coordinate lines, by its coordinate (hyper)surfaces or by a convenient set of these two ingredients. But there are obstructions to construct generic location systems by means of their parameterized (congruences of) lines. These obstructions are in part of structural or mathematical character, and in part of physical character. On one hand, in order that four congruences of lines be able to be parameterized in such a way that they constitute the coordinate lines of a coordinate system, they must obey
constraint equations which are in general incompatible for generic congruences. On the other hand, even if one can imagine a dust of micro-clocks as one congruence of lines, four of such congruences will impose serious, in general insoluble, problems of time scale, synchronization and individual accelerations of the clocks, the problem being more serious for light beams \cite{TolocoorLuz}. In short, only in particular space-times and under particular conditions location systems may be constructed by means of their coordinate lines. Consequently, the physical fields able to construct generic location systems are those defining one-parameter families of hypersurfaces. This is because four one-parameter families of hypersurfaces constitute generically the coordinate hypersurfaces of a coordinate system. Now, one clock broadcasting its proper time describes in the space-time a time-like line of which every event is the vertex of the future light cone formed by the electromagnetic signal broadcasting the time of the event, so that the set of these cones constitute a one parametric (proper time) family of (null hyper)surfaces. So, the four clocks broadcasting their proper time construct  physically the coordinate hypersurfaces of a coordinate system.

At every event in the domain of such a coordinate system, a receiver able to read the value of the proper time coded by every one of the four cones containing the event will obtain the four times $\{\tau^1,\tau^2,\tau^3, \tau^4\}$ that constitute the coordinates of the event.

What about the coordinate lines of such a positioning system? The coordinate lines of a coordinate system are the locus of points where all but one of the coordinate hypersurfaces cut together. A light cone contains either light-like  or space-like directions, the first ones being the generatrix, so that as a consequence, the intersection of three non tangent light cones cannot be but a space-like curve. Consequently, the coordinate lines of a positioning system constitute four parameterized  congruences of space-like lines.

As we see, in spite of the fact that they are physically well defined, positioning systems constitute the physical realizations of coordinate systems which are unusual for us. Thus, meanwhile we are accustomed to coordinate systems on which a point-like object may have, along its evolution, up to three constant (adapted) coordinates, the four emission coordinates of a positioning system necessarily change whatever the point-like material object be.  Nevertheless, the incomparable  operational character of such systems is worthy of an effort to better understand them and, on the way, to liberate ourselves of unjustified  Newtonian prejudices about the space-time.

\vspace{0.5cm}

In a {\em grid} of parameters $\{\tau^1,\tau^2,\tau^3,\tau^4\}$, any user  receiving continuously his coordinates may draw his trajectory.
An important class of positioning systems are the {\em auto-located positioning systems}, which allow the user to know also the trajectories of the emitting clocks in the grid.

The necessary and sufficient condition for a positioning system to be auto-located is that every clock broadcast the proper time that it directly receives from the other clocks.

This is because, joint to its proper time, the three times that the clock receives constitute in fact its proper coordinates. More precisely, the clocks are at the border of the coordinate domain that they generate, because one of the light cone coordinate hypersurfaces  of the domain is not differentiable at the positions of every clock, but the coordinates themselves are continuous along the world line of the clocks.

In a auto-located positioning system, a user receives at every instant sixteen times, $\{\tau^{ji}\},$ where $\tau^{ii} \equiv \tau^i$ is
the proper time of the clock $i,$  and
$\tau^{ji},$ $i \neq j,$ are the times received by the clock $i,$ from
 the clocks $j.$ Then, $\{\tau^i\}$ are the coordinates of the user and $\{\tau^{ji}\},$ $\forall j,$ are the coordinates of the clock $i.$

\vspace{0.5cm}

    The simplest examples of positioning system are found in a two-dimensional space-time, as shown in  Figure 1(a). In the internal region delimited by the trajectories of two clocks, the positioning system, constituted by the radiated electromagnetic fields broadcasting  the proper time of the clocks, is well defined (the exterior regions correspond to the shadow  of every clock  for the signal of the other, and are characterized by a vanishing Jacobian).
The emission coordinates of a user are the data     $(\tau^1\tau^2)$ that he receives.

    Because of the linearity of the light cones, the analysis of the two-dimensional case is particularly easy, and allows to understand with no much effort interesting features which, for the most part, remain qualitatively valid in higher dimensions. Nevertheless, dimension two is  singular for some properties, in part due to the fact that coordinate lines and coordinate surfaces coincide.

    Figure 1(b) shows an auto-located two-dimensional positioning system. Here, every user receives the four data $(\tau^1, \tau^{12}; \tau^2, \tau^{21})$ from which he can extracts his proper emission coodinates $(\tau^1, \tau^2)$ and also the emission coordinates $(\tau^1, \tau^{12})$ and $(\tau^{21}, \tau^2)$ of the satellites 1 and 2 respectively.

    Some basic properties of the two-dimensional case are presented in this same meeting by Ferrando \cite{Fe:2005}.

\begin{figure}
  \includegraphics[height=.25\textheight]{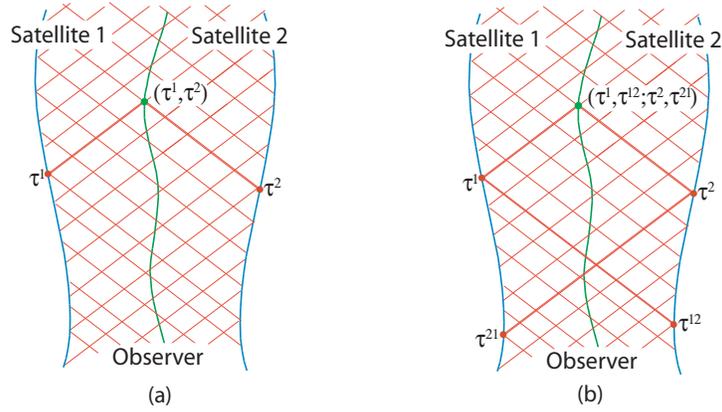}
   \caption{(a) Two-dimensional positioning system. (b) Auto-located two-dimensional positioning system.}
\end{figure}

An important property of positioning systems is that, for a given precision of the clocks,  they  improve the separation power of the events with respect to Cartesian location systems. The definition of the meter being based in that of the second, we can consider a Cartesian location system as constituted, at best, by four clocks of a given {\em precision}, i.e. of a given {\em separation power} of instants, one of them devoted to the measure of the time coordinate of the events, and the other three assigned to the construction of the three Cartesian space-like distance coordinates of them. The interesting result is that, if {\em same} precision clocks are assigned for the construction of emission coordinates of a positioning system, there always exists some space-time regions where the separation power of space-time events in these emission coordinates is  three times greater than that obtained by a Cartesian  protocol.

Other interesting  basic properties of positioning systems in the four-dimensional case  are presented in this same meeting by Pozo \cite{Po:2005}.

\section{SYPOR project}

The Global Positioning System (GPS) is a classical positioning system based on a set of satellites    around the Earth, that allows their users, receiving the signals of some of the satellites, to know their positions with respect to the Earth surface (nominally in the Word Geodetic System, WGS 84).
The information about the trajectory of the satellites with respect to the WGS 84 is centralized in the Master Control Monitory Station, at Colorado Springs, where the equations for the predicted trajectories are elaborated and sent to every satellite. Other GNSS, actual or in progress (GLONASS, Galileo) are based in a similar structure.

The GPS is a jewel of the military engineering, has largely fulfilled the schedule of conditions specifying delivering time and performances and offers an almost unlimited area of civil applications. It is clear that:
\begin{itemize}
\item[-] as a {\em technological object}, of engineering interest, the only possible improvements seem to be those derived from the technical improvements of its components, but
\item[-] as a {\em physical object}, of scientific interest, it seems nevertheless to admit radical conceptual innovations...
\end{itemize}

The need for these conceptual innovations is based in the following facts. It starts with an incorrect theory, the Newtonian one, that must be corrected with 'relativistic terms'. From the beginning, it constraints times and synchronizations with respect to the GPS time, a sort of advanced TAI (International Atomic Time), which is a global conventional time, not a local physical one. It uses the satellites not as the best supports for standard clocks (microgravity), but as (unfortunately moving!) beacons controlled from the Earth. Control and positioning are made with respect to the WGS 84, a reference system (not a positioning one!) which, in addition, is a virtual one (see above).

\vspace{0.5cm}

The aim of the project SYPOR (SYst\`emes de POsitionnement Relativistes)is to construct a {\em complete relativistic theory} of GNSS, i.e. a theory in which {\em only relativistic concepts} are involved, irrespective of the acceptable {\em numerical} simplifications that error bars and weakness of some quantities can justify. Such a theory will be {\em convergent} for increasing precision, meanwhile the present one is clearly {\em divergent}, and strongly mixed in its basis with Newtonian conventional protocols.

It is true that, for purposes of geodesy and positioning, the Earth may be frequently considered as a Newtonian system, i.e. a physical system correctly described by Newtonian theory. But a constellation of satellite-borne clocks interchanging their proper time around the Earth is a {\em relativistic system} on its own (principally because the importance of the gravitational and Doppler  correction terms). Consequently, the best, shortest and clearest way to improve present GNSS is to directly use the best concepts of relativity theory.

For this reason, the project SYPOR proposes to uncouple GNSS in two hierarchical systems:

\begin{itemize}
    \item  A primary system, Earth-surface independent, constituted by the constellation of satellites acting as an atlas (union of sets of four neighboring satellites) of primary auto-located relativistic positioning systems, related only to the mass content of the Earth. Its physical realization implies an Inter Satellite Link (ISL) between neighboring satellites, a device on every satellite to send to the Earth the links directly received by every satellite, and a device on some of the satellites of the system to connect the constellation of satellites to the orientation system ICRS.

    \item A secondary system, Earth-surface dependent, coupling the virtual and intrinsically retarded Earth reference system (WGS 84 or ITRF)  to the real and immediate primary main system.
\end{itemize}

A {\em space agency} could, or even should, limit its task to realize the
 primary system, and delegate to global or local Earth agencies the task of attaching (secondary) terrestrial reference systems to it.

Let us remark that if the primary system alone does not allow the users to situate with respect to the Earth surface, it nevertheless allow every user to situate with respect to the constellation (or with respect to any more conventional reference system {\em deduced} from it) and also allow  two  or more users to know their relative positions.

\vspace{0.5cm}

A final remark. The TAI may be improved by satellited clocks, as is contemplate by the project ACES (Atomic Clock Ensemble in Space). But with the notion of relativistic position system in mind, it becomes clear that four or more satellited atomic clocks not only are able to supply an International Atomic Time, but also to constitute an International Atomic Coordinate System (SCAI).

 Such a system would reconcile people more deeply and faithfully with relativity theory than the most part of the relativistic chattering  that we have contemplate this year all over the Earth.

%%%%%%%%%%%%%%%%%%%%%%%%%%%%%%%%%%%%%%%%%%%%%%%%
%% BACKMATTER
%%%%%%%%%%%%%%%%%%%%%%%%%%%%%%%%%%%%%%%%%%%%%%%%

\begin{theacknowledgments}
  Long time ago, this subject covered a corner of my private garden of thoughts for weekends and holidays.  But  every flower that sprouted in it, every idea, I showed it to my friends Joan F{\footnotesize{ERRANDO}}, Juan Antonio M{\footnotesize{ORALES}}, Albert T{\footnotesize{ARANTOLA}} and Jos\'e Maria P{\footnotesize{OZO}}, who watered it carefully. For this reason, it is a pleasure for me to acknowledge their important contribution to the subject.
\end{theacknowledgments}

%%%%%%%%%%%%%%%%%%%%%%%%%%%%%%%%%%%%%%%%%%%%%%%%
%% The bibliography can be prepared using the BibTeX program or
%% manually.
%%%%%%%%%%%%%%%%%%%%%%%%%%%%%%%%%%%%%%%%%%%%%%%%

%%%%%%%%%%%%%%%%%%%%%%%%%%%%%%%%%%%%%%%%%%%
%% The following lines show an example how to produce a bibliography
%% without the help of the BibTeX program. This could be used instead
%% of the above.
%%%%%%%%%%%%%%%%%%%%%%%%%%%%%%%%%%%%%%%%%%%


\begin{thebibliography}{9}

\bibitem{Fe:2005}
J.~J.~Ferrando, ``Coll positioning  systems: a two-dimensional approach,'' in  \emph{these proceedings} (2005).

\bibitem{Mo:2005}
J.~A.~Morales, ``Coordinates and frames from the causal point of view,'' in \emph{these proceedings} (2005).

\bibitem{Po:2005}
J.~M.~Pozo, ``Emission coordinates and the central observer,'' in \emph{these proceedings} (2005).

\bibitem{CT}
B.~Coll and A.~Tarantola,  ``A Galactic positioning system,'' in
  \emph{Journ\'ees Syst\`emes de R\'ef\'erence Spatio-Temporels},
  St. Petersburg, 22-25 September (2003), edited by A. Finkelstein and N. Capitaine, I.A.A., Russian Academy of Sciences and Observatoire de Paris, 2004, p 333. See also http://coll.cc\,.

\bibitem{TolocoorLuz}
B.~Coll,  ``Light coordinates in relativity,'' in
  \emph{Spanish Relativity Meeting ERE 85}, edited by Pub. Servei de Publications de l'ETSEIB, Barcelona, 1985, p 29 (Spanish text). See http://coll.cc for an English translation.

\end{thebibliography}
\end{document}